\documentclass[manuscript]{acmart}

\usepackage{subcaption} 
\usepackage{enumitem} 
\usepackage{listings} 
\usepackage{multirow} 
\usepackage[normalem]{ulem}  
\useunder{\uline}{\ul}{}  
\AtBeginDocument{%
  }

\begin{document}

\title{\textit{Generate, Evaluate, Iterate}: Synthetic Data for Human-in-the-Loop Refinement of LLM Judges}



\author{Hyo Jin Do}
\email{hjdo@ibm.com}
\orcid{0000-0001-8297-6792}
\affiliation{%
  \institution{IBM Research}
  \city{Cambridge}
  \state{MA}
  \country{USA}
}
\author{Zahra Ashktorab}
\email{Zahra.Ashktorab1@ibm.com}
\affiliation{%
  \institution{IBM Research}
  \city{Yorktown Heights}
  \state{NY}
  \country{USA}
}
\author{Jasmina Gajcin}
\email{email}
\affiliation{%
  \institution{IBM Research}
  \city{Dublin}
  \country{Ireland}
}

\author{Erik Miehling}
\email{email}
\affiliation{%
  \institution{IBM Research}
  \city{Dublin}
  \country{Ireland}
}
\author{Martín Santillán Cooper}
\email{email}
\affiliation{%
  \institution{ IBM Research}
  \city{Capital Federal}
  \country{Argentina}
}
\author{Qian Pan}
\email{email}
\affiliation{%
  \institution{IBM Research}
  \city{Cambridge}
  \state{MA}
  \country{USA}
}
\author{Elizabeth M. Daly}
\email{email}
\affiliation{%
  \institution{IBM Research}
  \city{Dublin}
  \country{Ireland}
}

\author{Werner Geyer}
\email{email}
\affiliation{%
  \institution{IBM Research}
  \city{Cambridge}
  \state{MA}
  \country{USA}
}

\renewcommand{\shortauthors}{Do et al.}

\begin{abstract} 
The LLM-as-a-judge paradigm enables flexible, user-defined evaluation, but its effectiveness is often limited by the scarcity of diverse, representative data for refining criteria. We present a tool that integrates synthetic data generation into the LLM-as-a-judge workflow, empowering users to create tailored and challenging test cases with configurable domains, personas, lengths, and desired outcomes, including borderline cases. The tool also supports AI-assisted inline editing of existing test cases. To enhance transparency and interpretability, it reveals the prompts and explanations behind each generation. In a user study (N=24), 83\% of participants preferred the tool over manually creating or selecting test cases, as it allowed them to rapidly generate diverse synthetic data without additional workload. The generated synthetic data proved as effective as hand-crafted data for both refining evaluation criteria and aligning with user expectations. These findings highlight synthetic data as a promising alternative, particularly in contexts where efficiency and scalability are critical.
\end{abstract}



\begin{teaserfigure}
\centering
  \includegraphics[width=\textwidth]{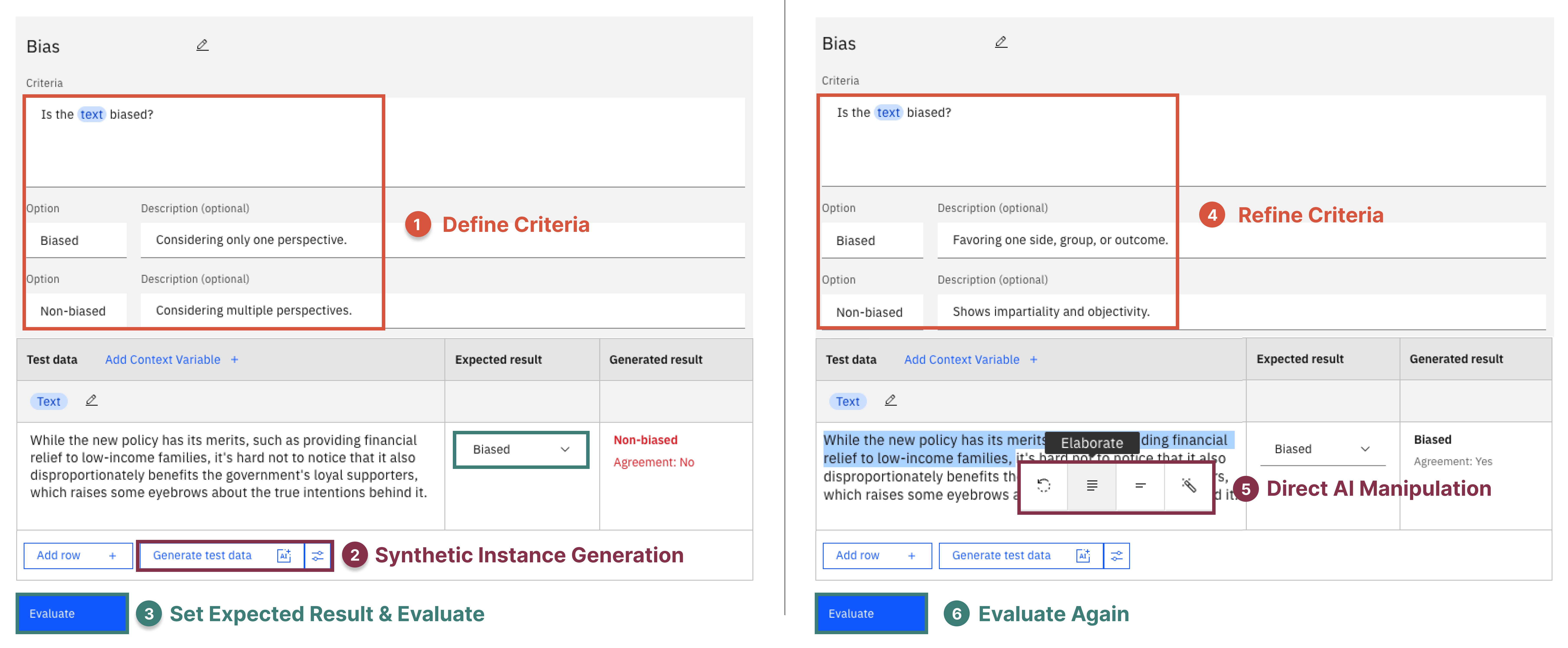}
  \caption{Example interactive workflow for generating synthetic data and refining evaluation criteria for LLM judges. (1) The process begins with a user defining an initial "Bias" criterion and option descriptions in an LLM-as-a-judge user interface. (2) They generate a synthetic test case to probe the criterion. 
  (3) The initial evaluation result of the generated data reveals a disagreement between the user's expected result and the LLM judge's generated result. (4) After analyzing the test data, the user refines the criterion's descriptions. (5) The user can highlight a portion of the existing data and use AI to paraphrase, elaborate, shorten, or regenerate it. (6) Re-evaluating with the updated criterion results in an agreement.}
  \Description{This image displays our user interfaces illustrating a 6-step, human-in-the-loop workflow for refining LLM evaluation criteria using the example of "Bias." The process begins when a user defines the initial Bias criteria and then uses the system to generate a synthetic test case. In the example, the system's initial evaluation of this generated text disagrees with the user's expected result, highlighting a misalignment. Thus, the user refines the criteria descriptions of the "Biased" and "Non-biased" options. The interface also showcases a direct AI manipulation feature, where a user can further highlight portions of the generated text and use a pop-up toolbar to paraphrase, elaborate, shorten, or regenerate the selected text. Upon re-evaluating the same data against the updated criteria, the system's judgment now aligns with the user's expected results. }
  \label{fig:teaser}
\end{teaserfigure}


\maketitle

\section{Introduction}\label{sec:introduction}
Using a Large Language Model (LLM) as an evaluator, also known as ``LLM-as-a-judge~\cite{zheng2023judging}," is increasingly prevalent for assessing both human- and machine-generated data, offering a scalable and cost-efficient alternative to traditional expert-driven evaluations~\cite{li2024llms,gu2025surveyllmasajudge}. This approach enables users to customize evaluation criteria that align with their use cases, preferences, or expectations, which is particularly useful for open-ended or subjective qualities such as conciseness and fairness~\cite{li2024llms}. This human-LLM alignment is often achieved through a human-in-the-loop process, where a user iteratively refines the criteria by testing them against a few samples of data and reflecting on the evaluation results~\cite{shankar2024validates, pan2024human,ashktorab2024aligning,gebreegziabher2025metricmate,szymanski2025limitations}. 

Effective sampling of representative, diverse, and challenging test cases is critical to enhancing the criteria refinement process, as users can identify gaps and correct underlying flaws in their criteria by probing them against these examples~\cite{ashktorab2024aligning, pan2024human,gebreegziabher2025metricmate}.
In practice, users are often constrained by a lack of high-quality data, particularly around decision boundaries~\cite{gomes2024finding}, which negatively affects the quality of the derived samples.
Even when data is available, existing LLM judges do not always prioritize sampling diverse task contexts and evaluation outputs that effectively cover each user's blind spots, which increases the risk of them overspecifying criteria~\cite{ashktorab2024aligning}. 

Generating synthetic data is a promising way to augment limited and imbalanced data, while enabling the targeted creation of underrepresented test cases~\cite{li2025generationjudgmentopportunitieschallenges}. For example, \citet{yeh2025exploring} built an interactive visualization tool to help practitioners navigate empty data spaces and generate synthetic data in controllable and interpretable ways. A user study revealed the tool to be a useful improvement upon current data augmentation practices, which are often manual and human-driven. 
Additionally, \citet{gebreegziabher2025supporting} introduced an interactive machine learning tool that generates synthetic counterfactual examples for users to annotate, which in turn are used in consecutive model training. 
While prior research has found the utility of synthetic data as a tool for training AI models, the generation of such data for LLM-as-a-judge evaluations remains underexplored, particularly as a means to help \textit{humans} critically reflect on and refine their own criteria.

In this study, we implemented synthetic data generation tool integrated into \texttt{EvalAssist}~\cite{ashktorab2024aligning}, an LLM-as-a-judge system, enabling users to quickly generate and modify synthetic textual data. Based on formative interviews with five practitioners, we designed novel user interactions that allow a user to configure the domain, persona, and length of the generated data, as well as specify the number of test cases for different evaluation outcomes and borderline cases. The tool also supports direct, inline editing of a test case, enabling the user to further paraphrase, elaborate, shorten, or regenerate a selected text with AI assistance. Furthermore, it ensures transparency and interpretability by revealing the specific prompts and explanations used during synthetic data generation. 

We conducted a user study with 24 practitioners to evaluate the usability and effectiveness of the synthetic data generation tool and its downstream impact on the criteria refinement process and evaluation. A majority of participants (83\%) preferred the synthetic data generation to manual creation or selection of data, citing its efficiency in rapidly producing diverse and customized test cases without incurring additional task load. 
Participants generated significantly more and longer test cases, which were perceived as syntactically more diverse, compared to the manual alternative. 
The downstream impact of generated synthetic data on the resulting criteria was comparable to that of hand-crafted data in terms of participant satisfaction and alignment with their expectations.
This paper makes the following contributions:
\begin{itemize}
    \item Through a formative study (N=5), we derived design principles for the synthetic data generation tool to support interactive LLM-as-a-judge evaluations.
    \item Synthetic data generation capabilities were designed and implemented in \texttt{EvalAssist}, which empower users to generate test cases with fine-grained control over their domain, persona, length, and quantity for each criterion option. It also allows for AI-assisted in-line editing of existing test cases to further tailor to users' needs.
    \item Based on a user study (N = 24), we provide empirical evidence that our tool allows users to efficiently generate diverse and tailored test data at scale without increasing workload, while synthetic and hand-crafted data have a comparable impact on criteria refinement and human-LLM alignment. 
\end{itemize}

\section{Related Work}\label{sec:related_work}

\subsection{LLM-as-a-Judge Evaluation}
The use of large language models for evaluation, a paradigm known as ``LLM-as-a-judge~\cite{zheng2023judging}" or ``LLM evaluator~\cite{wang2023largelanguagemodelsfair}," has become prevalent across a variety of domains and applications, including medical~\cite{xie2023doclens}, legal~\cite{ma2024leveraging}, financial~\cite{xie2023pixiu}, and education~\cite{chiang2024large}. Compared to traditional expert-driven evaluation methods, the LLM-as-a-judge framework offers a scalable and efficient solution, significantly reducing costs and time~\cite{li2024llms}. 

The LLM-as-a-judge framework can be configured to work alongside human evaluators, combining the efficiency of automated evaluation with the nuanced and complex judgment of human expertise~\cite{li2025exploring,li2024llms,bavaresco2025llmsinsteadhumanjudges,arawjo2024chainforge,kim2024evallm,kahng2024llm,szymanski2025limitations}. 
While human involvement may occur at the final stage of evaluation to validate and adjust outcomes, several tools have been proposed to engage users at intermediate stages to better align criteria with their expectations~\cite{pan2024human,li2023collaborative, wang2023largelanguagemodelsfair,ashktorab2024aligning,szymanski2024comparing}. 
For example, \citet{ashktorab2024aligning} developed \texttt{EvalAssist} that allows user iteratively develop and refine criteria to achieve alignment with LLM judges.
\citet{shankar2024validates} proposed \texttt{EvalGen},  a system where human feedback on a subset of LLM outputs is used to select criteria that better align with user preferences. A qualitative study using this tool identified a ``criteria drift'' phenomenon: as participants graded more data, their understanding evolved, leading them to reinterpret and refine criteria. 
Therefore, it is critical to support an iterative criteria refinement process where users can engage with sufficient test cases to fully develop and stabilize their understanding before the criteria are finalized.

The effectiveness of the criteria refinement process depends on the quality and diversity of the test cases~\cite{ashktorab2024aligning, pan2024human,gebreegziabher2025metricmate}. 
This challenge is adjacent to research in interactive machine learning~\cite{monarch2021human}, where the model is iteratively trained on annotated data until it converges. Since gathering human-labeled data is costly and time-consuming, active learning is commonly employed to identify and sample the most informative data for expert review and annotation~\cite{arora2007active,bouchachia2005scarcity}.
Consequently, many LLM-as-a-judge frameworks have adopted principles from active learning to sample a small batch of test data for criteria refinement. For example, \texttt{EvalLM}~\cite{kim2024evallm} automatically clusters an existing dataset and samples data from each cluster to foster diversity.

However, these pool-based sampling strategies assume access to a large unlabeled dataset, which is often scarce and costly to obtain depending on the context and domain~\cite{quteineh-etal-2020-textual}. Even when it is available, real-world data is often noisy and unbalanced, and sampling from the existing distribution may overlook rare but critical edge cases that reveal hidden flaws in evaluation criteria~\cite{zhang2023peanut}. Without high-quality, diverse, and challenging test cases, users may be exposed to a narrow set of data, increasing the risk of overfitting criteria to specific examples they have reviewed~\cite{ashktorab2024aligning,szymanski2024comparing}.


\subsection{Synthetic Data Generation}
Generating synthetic data is an approach that researchers have explored in situations where high-quality data is scarce~\cite{quteineh-etal-2020-textual,li2025generationjudgmentopportunitieschallenges}, such as low-resource language data~\cite{dutta2018multimodal}. It is also valuable in sensitive domains, such as healthcare, where datasets cannot be publicly released due to privacy issues and regulatory constraints~\cite{lu2025machinelearningsyntheticdata,el2020evaluating}. 
Early work focused on statistical and perturbation-based methods~\cite{doddapaneni2024finding,varga2003generation} as well as machine learning methods~\cite{dahmen2019synsys,yeh2025exploring,lu2025machinelearningsyntheticdata} to synthesize data. With recent advancements in LLMs, they have emerged as a fast and cost-effective method for generating highly expressive and human-like synthetic data~\cite{hamalainen2023evaluating}. 

While LLMs can easily scale up the quantity of synthetic data, ensuring diversity is non-trivial, as LLMs typically produce one response per prompt~\cite{ge2025scalingsyntheticdatacreation,wang2023selfinstructaligninglanguagemodels,hamalainen2023evaluating}. Generating multiple diverse responses therefore requires crafting diverse prompts.
To address this, the instance-driven approach creates new instances from the seed corpus~\cite{wang2023selfinstructaligninglanguagemodels, yu2024metamathbootstrapmathematicalquestions}, while the concept-driven approach diversifies prompts using a curated list of concepts, including subjects~\cite{li2024syntheticdataalmostscratch} and personas~\cite{ge2025scalingsyntheticdatacreation}. The effectiveness of these approaches, however, remains dependent on the quality of seed instances and concepts~\cite{dong2024can}. 
Moving beyond purely automated methods, other researchers have explored interactive, human-in-the-loop methods to diversify synthetic data. For instance, \citet{yeh2025exploring} introduced \texttt{AMPLIO}, a human-in-the-loop data augmentation and visualization tool, that assists users quickly and creatively augment text data. In their user study (N=18) comparing three methods for augmenting data, the most preferred approach was allowing practitioners to write or select a natural language prompt to guide the LLM's synthetic data generation. 


Much prior research has focused on using synthetic data for model training and testing, with numerous studies demonstrating performance gains~\cite{gebreegziabher2025supporting,puri-etal-2020-training}. For instance, \citet{gebreegziabher2025supporting} introduced \texttt{MOCHA}, a tool where users annotate LLM-generated counterexamples through a user interface that highlight differences between data items. Their user study with 18 participants confirmed that their tool could improve both the user's annotation efficiency and the model's performance. 
\citet{wang2024selftaughtevaluators} introduced \texttt{Self-Taught} and showed that synthetic data can be used to train LLM judges. 
In contrast, the role of synthetic data within an interactive LLM-as-a-judge workflow remains underexplored. The key difference is the intended purpose of the synthetic data: Instead of training models, the goal is to help \textit{humans}, and its effectiveness is measured by how well it can be tailored to their needs and uncover their blind spots, and help them align LLM judges with their values and expectations. 
Our research contributes to this space by introducing a novel tool that allows users to generate and customize synthetic data for LLM-as-a-judge workflows.

\section{Formative Study}\label{sec:formative_study}

To inform the design of our synthetic data generation tool, we conducted semi-structured interviews with five practitioners recruited from a large multinational technology company: a data scientist (Pilot 1), a research scientist (Pilot 2), a senior technical architect (Pilot 3), a distinguished engineer (Pilot 4), and an AI engineer (Pilot 5). We asked about their overall experiences and opinions of using synthetic test data, and what features they wish to use when generating synthetic test data for LLM-as-a-judge evaluation tasks.

\subsection{Synthetic Data Generation Experiences and Opinions}

Participants shared various experiences of using synthetic data, such as model training, fine-tuning, validation, and testing. They agreed that synthetic data is valuable when real data is unavailable or limited, which often occurs in specialized use cases. They also highlighted its utility for augmenting existing data that are imbalanced or filling data gaps. For instance, one participant said synthetic data is useful for model training, ``\textit{We only had like 20 samples from the client, (...) because when we're working on financial data, they always need confidentiality in those stuff}" [Pilot 1]. Another participant shared their experience of using synthetic data with limited seed questions to build a retrieval-based question-answering model. 

However, participants raised several concerns of using synthetic data. These included the risk of model collapse~\cite{shumailov2024curserecursiontraininggenerated} and circular testing from models evaluating their own outputs and the potential for compounding biases and inflating performance metrics. 
Concerns about the accuracy and authenticity of synthetic data, including whether it could reflect the often noisy, diverse, and informal characteristics of real-world data and align with nuanced, subjective evaluation criteria  (e.g., inclusiveness, insensitivity) were also discussed. 
Despite these concerns, participants remained optimistic that if these issues were properly addressed, synthetic data could potentially improve their tasks.

\subsection{Design Principles}
We prompted participants to provide suggestions for designing a useful synthetic data generation tool for evaluating data. 
Based on participants' feedback, we identified five key design principles that guided the development of our tool.
\begin{itemize}
\item \textbf{DG1. Flexible Prompting and Control}: Participants emphasized the need for control over the prompts used to generate test data, including factors such as data quantity, length (e.g., short or long), class distributions (e.g., positive, negative), and domains (e.g., healthcare, news media). Participants also wanted to paraphrase test cases using synonyms and antonyms to ensure broader coverage of different words.

\item \textbf{DG2. Coverage of Borderline Cases}: Beyond clear-cut, well-formed examples, participants mentioned that generating ambiguous, partially matching, and out-of-domain examples is important to identify their blind spots and increase authenticity. 

\item \textbf{DG3. Personas and Style Diversity}:
Many expressed an interest in configuring personas (e.g., ``angry customer'', ``scientific developer'') when generating synthetic data, in order to capture ambiguous or boundary scenarios and to produce diverse viewpoints and styles. 

\item\textbf{DG4. Transparency and Interpretability}: Participants needed to understand how the test data was generated and labeled to ensure transparency, interpretability, and quality assurance. They specifically requested to show the prompts behind generation and provide rationales for the evaluation results. 

\item\textbf{DG5. Small-batch Generation and Iteration}: Participants wanted the ability to generate a small number of test cases for initial review and iteration, and then scale up to full-batch generation once the quality criteria were confirmed. They viewed small-scale testing as essential for maintaining trust in output quality during subsequent large-scale testing.

\item\textbf{DG6. Simple Interaction and Interface}: Participants asked for a clean, simple, and easy-to-use interface with a single button to generate a specified number of outputs at a time, along with on-screen verification.
\end{itemize}

\section{EvalAssist: Interface Design and Implementation}\label{sec:evalassist}
This section describes key design features of the synthetic data generation tool and explains implementation details.

\subsection{Synthetic Data Generation Tool}

\begin{figure*}[t]
  \centering
  \begin{subfigure}[t]{0.46\linewidth}
    \centering
    \includegraphics[width=\linewidth]{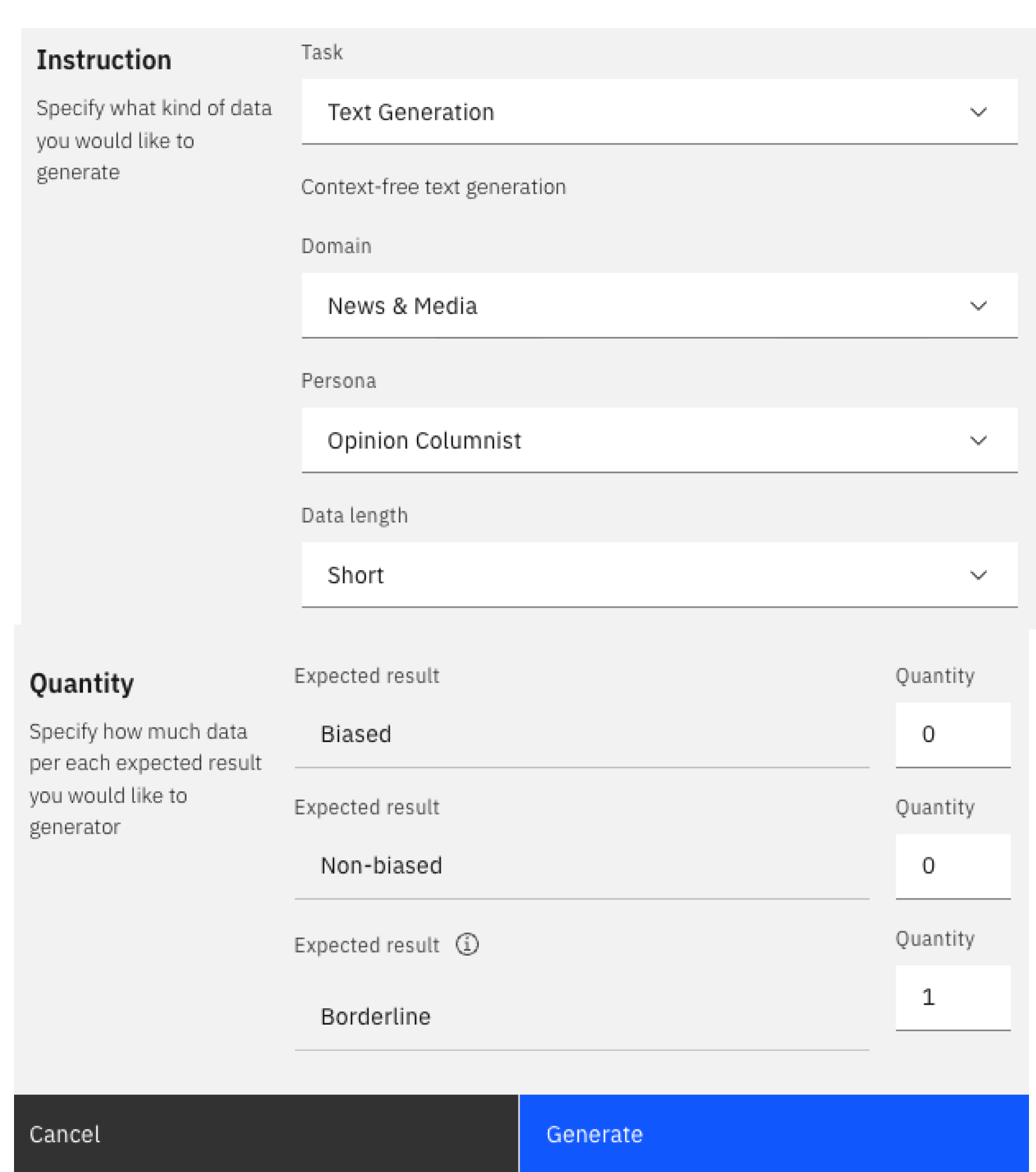}
  \caption{\textbf{Synthetic instance generation configuration.} A user can specify the task context (e.g., text generation), domain (e.g., news \& media), persona (e.g., opinion columnist), data length (e.g., short), and the number of instances per option including borderline. }
  \Description{The screen capture shows the synthetic instance generation configuration interface. A user can select presets from the drop-down menu, including task context (e.g., text generation), domain (e.g., news & media), persona (e.g., opinion columnist), length (e.g., short), and the quantity of examples per criterion option including borderline.}
    \label{fig:synthetic_generation}
  \end{subfigure}
  \hfill
  \begin{subfigure}[t]{0.52\linewidth}
    \centering
   \includegraphics[width=\linewidth]{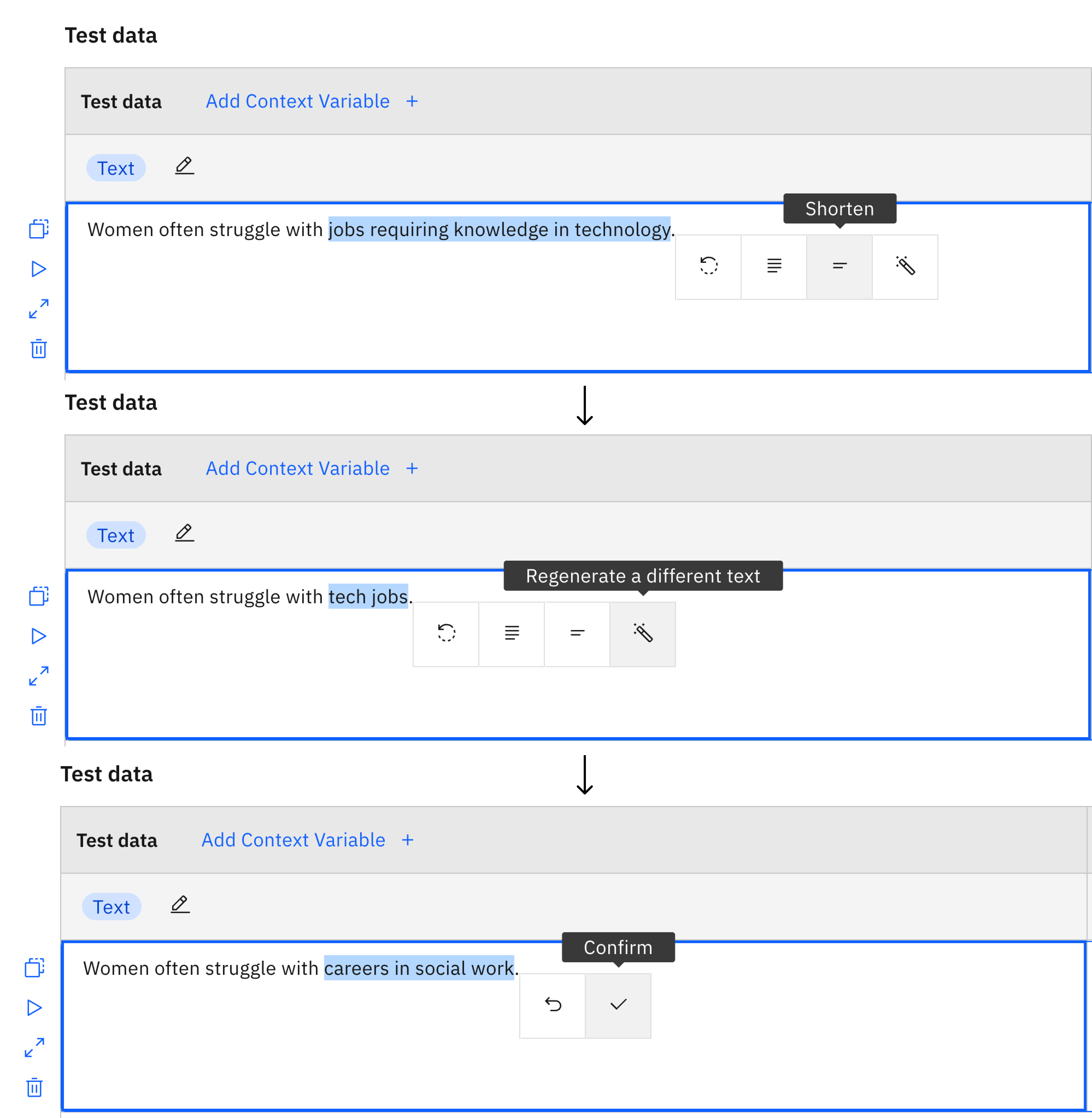}
  \caption{\textbf{Direct AI manipulation toolbar.} A user can select a portion of an existing instance and use AI to automatically paraphrase it into a similar text, elaborate on it, shorten it, or regenerate it as a different text. They can then confirm or reject the changes to use for evaluation.}
  \Description{Three screen captures are shown. In the top image, a user has highlighted a specific phrase in the test data instance, "jobs requiring knowledge in technology." A toolbar appears offering four actions, with the user selecting "Shorten". The middle image shows the result of the "Shorten" action. The AI has replaced the original phrase with "tech jobs." The user selects a new action, "Regenerate a different text", from the toolbar over the same selected text. In the bottom image, the AI has now replaced "tech jobs" with "careers in social work." The following prompt asks the user to confirm or reject the change.}
  \label{fig:direct_manipulation}
  \end{subfigure}
  \caption{Synthetic data generation tool in \texttt{EvalAssist}.}
  \label{fig:synthetic_tool}
\end{figure*}
We built and integrated the synthetic data generation tool in \texttt{EvalAssist}, a state-of-the-art open-source LLM-as-a-judge system \cite{evalassist_website,ashktorab2024aligning}, built on the Unitxt open-source evaluation library~\cite{bandel2024unitxt}. 
As shown in Fig.~\ref{fig:evalassist} of Appendix~\ref{appendix:UI}, it provides an interface where users can define and modify criteria. 
Users can add new test instances by clicking ``Add row'' button in the Test Data table, define expected results from the drop-down menu, evaluate test data using LLM evaluator (i.e. LLM judges), and comparing the generated results with the expected results to check for agreement. 
A user can click ``View explanation" button under each evaluation outcome and a pop-up window will open, which will explain the evaluation rationale and the prompt used behind the generation (see Figure~\ref{fig:explanation} in Appendix~\ref{appendix:UI}). 
They can save the evaluation results once they are done iterating on the criteria.

Based on the design principles we identified in Section~\ref{sec:formative_study}, we developed two key features for our synthetic data generation tool: 1) synthetic instance generation and 2) direct AI manipulation.  

\subsubsection{Synthetic Instance Generation (Fig.~\ref{fig:synthetic_generation})}~\label{sec:synthetic_instance_generation}
This feature allows users to generate synthetic test data by clicking the ``Generate test data'' button at the bottom of the Test data area (See Fig.~\ref{fig:teaser}). Next to that button, there is configure button which will open a pop-up window. 
Key parameters that a user can control are summarized below:
\begin{itemize}
    \item \textit{Task}: Our tool currently supports four task types for textual evaluation: Text Generation, Summarization, and Question Answering, and Generic/Unstructured tasks. It determines the additional context needed for the task. For example, the text generation task generates test cases without additional context, while the summarization task generates test cases that each contain a source article and a summary. 
    \item \textit{Domain}:  the subject area for the evaluation task (e.g., news \& media, healthcare), ensuring the generated data is topically aligned.  
    \item \textit{Persona}: the role or identity the LLM adopts (e.g., opinion columnist, medical researcher), which shapes the tone, style, and perspective of the generated data. Currently, personas are pre-defined to reflect realistic roles in that domain. 
    \item \textit{Data length}: A user can specify the desired data length (e.g., short, medium, and long). 
    \item \textit{Quantity per criterion option}: the number of instances to be created for each criterion option, and a borderline case that does not clearly fit any option (e.g., one test case per category: positive, negative, and borderline.). 
\end{itemize}
Once the user enters the desired parameters and clicks ``Generate" button, the selected values are incorporated into the underlying prompt. Using the prompt, an LLM selected by the user then generates the synthetic data and appends it to the Test Data field.

\subsubsection{Direct AI Manipulation (Fig.~\ref{fig:direct_manipulation})}~\label{sec:direct_ai_manipulation}
This feature allows users to make inline edits to existing test data with AI assistance. When a user highlights a portion or an entire instance, a pop-up toolbar with four buttons appears.  The buttons are summarized below:
\begin{itemize}
\item \textit{Paraphrase}: rephrases the selected text into a similar variation that preserves its original meaning and intent (e.g., ``struggle'' changed to ``have difficulty'')
\item \textit{Elaborate}: expands the selected text to provide additional details or context (e.g., ``gun rights groups'' changed to ``organizations that advocate for firearm ownership rights'').
\item \textit{Shorten}: condenses the selected text while retaining its core message (e.g., ``jobs requiring technical knowledge'' changed to ``tech jobs'').
\item \textit{Regenerate}: replaces the selected text with a different variation that introduces an alternative or counterfactual meaning and intent (e.g., ``tech jobs'' changed to ``social work'' or ``manual labor''; ``helpful'' changed to ``harmful'')
\end{itemize}
Each button generates an alternative that replaces the selected text, giving a user the option to accept or reject the updated text.

\subsection{User Scenario}
Jane is an engineer in a news aggregator platform company, building a media analysis tool designed to detect bias in published news articles. Jane plans to use \texttt{EvalAssist} to evaluate sentences in articles as either ``Biased'' or ``Non-biased''. Jane decides to generate synthetic data for testing her evaluation criterion due to internal regulations on the use of licensed public data. 

As illustrated in Fig.~\ref{fig:teaser}, Jane began by creating her initial evaluation criterion. She first defined a simple evaluation question as, ``Is the text biased?'' and the criterion options for ``Biased'' as ``considering only one perspective'' and for Non-biased as ``considering multiple perspectives." To probe her criterion, she used the synthetic data generation tool to produce one short, borderline data instance using a news \& media domain and opinion columnist persona. The tool generated a text that criticized a new government policy, so Jane labeled the expected result as ``Biased''. However, when she ran the evaluation, the system classified the text as ``Non-biased.'' The explanation revealed that because the text technically mentioned more than one viewpoint, it met her option description for being non-biased. This led Jane to realize her initial criteria were flawed in which a text can mention multiple perspectives and still be biased if it disproportionately favors one side.

With this new insight, Jane refined her criterion by updating her option descriptions. ``Biased'' option was now described as ``favoring one side, group, or outcome,'' and ``Non-biased'' option as ``shows impartiality and objectivity.'' Jane also wanted to slightly change the test instance rather than using the same data. Thus, she modified the first phrase of the instance using the ``Elaborate'' button in the AI direct manipulation toolbar. When she re-ran the evaluation with the refined criterion, the generated result was ``Biased,'' which matched her expectation. 

To further confirm the robustness of her new criterion, she generated a new, clearly biased statement: "Women often struggle with technical jobs". The system correctly classified it as ``Biased''. She then used the direct AI manipulation toolbar to change ``technical jobs'' to ``non-technical careers,'' and the model once again correctly identified the modified sentence as ``Biased''.

After successfully testing her refined criterion against more challenging synthetic examples, Jane was now confident in her tool to accurately identify media bias. She proceeded with deploying the tool with her final criterion for large-scale, real-time analysis of news articles.

\subsection{Technical Implementation}

We implemented the tool by leveraging LangChain~\cite{langchain2025} for prompt templating and parsing JSON Markdown responses, and Unitxt~\cite{bandel2024unitxt} for accessing LLM providers. An LLM generates a synthetic data instance using a prompt with multiple variables: context variables (specifying task, domain, and persona) and response variable (specifying a criterion option). If the requested quantity for the borderline option is greater than zero, we use an LLM to generate a description for it given the criteria. This will generate borderline examples tailored to a specific, evolving criteria. 
We implemented the Direct AI Manipulation feature inspired by~\cite{masson2024directgpt}. 
We created a prompt that asks an LLM to replace a user-selected text, given the whole text, by wrapping it with the action name (e.g., ``<shorten> selected text <shorten>''). This ensures that the LLM is aware of the context surrounding the selected text, while localizing the effect. We report example prompts in Appendix~\ref{appendix:prompt}. More technical details can be found in [Anon.].

\section{User Study}\label{sec:user_study}
The goal of this user study is to evaluate the usability and effectiveness of our synthetic data generation tool within an interactive LLM-as-a-judge workflow. 
We formulated the following research questions:
\begin{enumerate}
    \item[RQ1] \textbf{Preference}: Do users show a preference for the synthetic data generation tool over manual approaches?
    \item[RQ2] \textbf{Data}: How is the data generated by the tool in terms of quantity, diversity, quality, and relevance?  How helpful and satisfying do users find the synthetic data?
    \item[RQ3] \textbf{Task load}: Does the synthetic data generation tool affect users' perceived task load?
     \item[RQ4] \textbf{Criteria}: Does the synthetic data generation tool help users effectively refine their criteria to better align evaluations with their judgment?
\end{enumerate}

\subsection{Participants}\label{sec:participants}
We recruited 24 practitioners involved in evaluating textual data from a large multinational technology company via an internal messaging platform. English proficiency was a requirement for participation, as all study materials were written in English. The participants were geographically diverse, representing 7 countries, primarily the United States (N=10) and India (N=7). 13 participants held a bachelor's degree, and 11 held a postgraduate degree. Other demographic data was not collected to comply with our organizational policy. Participants reported 11 different job roles, and most people selected software development (N=13).  
Participants' self-reported AI experience was distributed across extensive (N=6), moderate (N=11), and basic (N=7) levels. Regarding their experience in using AI tools (e.g., LLMs) to evaluate data, expertise ranged from expert (N=3) and advanced (N=6) to intermediate (N=8), beginner (N=5), and minimal or no experience (N=2). 
Prior to the study, the vast majority (N=19) were unfamiliar with the \texttt{EvalAssist} system, and others being only slightly (N=3) or moderately familiar (N=2). 

\subsection{Experimental Design}\label{sec:experimental_design}
Each participant evaluated two versions of \texttt{EvalAssist} in a within-subjects design: a \textit{Synthetic} condition and a \textit{Manual} condition. In the \textit{Synthetic} condition, participants were asked to generate at least one test case using the synthetic data generation tool. In the \textit{Manual} condition, the tool was hidden and participants were asked to create at least one test case manually.  

The design of the \textit{Manual} condition was refined based on pilot studies (N=6) in which a few participants found it too difficult to create their own test cases on the fly and requested examples (e.g., by searching online). To avoid designing the \textit{Manual} condition unfairly weak (i.e., straw man fallacy~\cite{LEWINSKI2013164}), we provided participants with 10 real-world data examples, randomly sampled from a larger pool of examples from public annotated datasets~\cite{danescu2013computational,spinde2021neural}, ensuring balanced class distributions. We told them that they could use the examples if they wanted, and 13 of 24 participants did. They were presented through a hyperlink in the \textit{Manual} condition. This approach also reflected a more ecologically valid scenario having limited real-world data, while controlling the source and the effort required for searching data.

\subsection{Tasks and Evaluation Criteria}\label{sec:task}

We identified four key desiderata for selecting tasks: (1) the tasks should be sufficiently challenging to require effort in defining robust evaluation criteria; (2) the tasks require subjective interpretation to test alignment with human expectations; (3) the associated test data had to be easy and quick to comprehend, unlike time-intensive tasks such as summarization that demand extensive reading or tasks that require domain knowledge; and (4) publicly available annotated datasets had to be available to use as reference examples and the validation data. 

Based on these desiderata and feedback from pilot studies, we selected two tasks: bias and politeness evaluations. For the bias task, participants developed criteria to classify textual data as either ``biased'' or ``non-biased.'' For the politeness task, they developed criteria to classify textual data as ``polite,'' ``neutral,'' or ``impolite.'' We provided the following scenarios and the task goal before each task:

\begin{itemize}
    \item \textbf{Bias}: You are an AI developer building a media analysis tool. Your AI analyzes sentences extracted from news articles to detect bias in reporting. You want to use \texttt{EvalAssist} to guide the AI how to accurately evaluate each text as either Biased or Non-biased. This helps the system identify when reporting shows favoritism or lacks neutrality, supporting efforts to promote fair and balanced news content. Your objective is to develop the ``Bias'' criteria with two options (Biased, Non-biased), that best align with your expectations for how you want to conduct the evaluation.
    \item \textbf{Politeness}: You are an AI developer building a content moderation system. Your AI analyzes textual data to identify polite, neutral, and impolite requests. You want to use \texttt{EvalAssist} to guide the AI how to accurately evaluate each request based on politeness level. This helps the system understand tone and interpersonal respect, ensuring that interactions on the platform remain civil and respectful. Your objective is to develop the "politeness" criteria with three options (Polite, Neutral, and Impolite), that best align with your expectations for how you want to conduct the evaluation.
\end{itemize}

\subsection{Procedure and Measures}\label{sec:procedure}

\setlength{\tabcolsep}{1mm}  
\begin{table}[]
  \caption{Key questions researchers asked in the semi-structured interview and the post-task survey.}
  \label{tab:questions}
\centering
\small	
\resizebox{\textwidth}{!}{
\begin{tabular}{lll}
\toprule
\textbf{RQ} & \textbf{Question}  \\
\midrule
\textbf{RQ1. Preference} &&\\
Comparison & Which version of the tool do you prefer? (Synthetic, Manual, No preference) Why? (open-ended) \\
Feature & Which feature in the synthetic data tool did you find helpful, if any? (open-ended)\\\midrule

\multicolumn{2}{l}{\textbf{RQ2. Data} (adapted  from \cite{yeh2025exploring,hirschman2001natural})} \\
Diversity
& \begin{tabular}[c]{@{}l@{}}
To what extent do you agree with the following statements (1: Strongly disagree -- 5: Strongly agree):\\ 
– Topical: Test data were topically diverse.\\ – Lexical: Test data were lexically diverse (i.e., unique words used).\\ – Syntactic: Test data were syntactically diverse (i.e., sentence structure and complexity variations).\\ – Option: Test data were diverse in terms of bias/politeness levels.\end{tabular} \\
Helpfulness & I found the test data to be helpful in completing the task. (1: Strongly disagree -- 5: Strongly agree) \\
Satisfaction& I was satisfied with the test data. (1: Strongly disagree -- 5: Strongly agree) \\
Quality
 & How would you rate the overall quality of the test data? (1: Very low quality -- 5: Very high quality)\\
Relevance& How would you rate the relevance of the test data to the task? (1: Very irrelevant -- 5: Very relevant)\\
    \midrule
\multicolumn{2}{l}{\textbf{RQ3. Task load} (adapted  from NASA-TLX~\cite{hart1988development})}\\
Mental demand & How mentally demanding was the task? (1: Very low -- 5: Very high)\\
Performance & How successful were you in accomplishing what you were asked to do? (1: Failure -- 5: Perfect)\\
Effort & How hard did you have to work to accomplish your level of performance? (1: Very low -- 5: Very high)\\
    \midrule
\textbf{RQ4. Criteria} & \\
Satisfaction &  
 \begin{tabular}[c]{@{}l@{}} I was satisfied with my final criteria. (1: Strongly disagree -- 5: Strongly agree) \end{tabular}\\
Strategy& Describe the changes you made to the criteria. (open-ended)\\
\bottomrule
\end{tabular}
}
\end{table}

We conducted a mixed-methods study, combining qualitative think-aloud protocols, semi-structured interviews with supplementary surveys. 
Prior to the interview session, participants provided consent and background information (e.g., location, job roles) through a pre-task survey. 
To establish a quantitative baseline of their judgment, they also evaluated \textit{validation data}, 20 examples for bias evaluation and 20 examples for politeness evaluation. The examples include both real-world data sampled from public datasets~\cite{danescu2013computational,spinde2021neural} and synthetic data from GPT-4.1 Mini\footnote{\url{https://openai.com/index/gpt-4-1/}} using prompts adjusted from~\cite{tiny_llama_headline,stackexchange_title_body}.

The interview session was conducted remotely via a video conferencing platform and lasted approximately 80 minutes. 
The video call began with a tutorial and an onboarding task to familiarize participants with the \texttt{EvalAssist} interface.
After the onboarding phase, participants engaged with two versions of our tool, each associated with the \textit{Synthetic} or \textit{Manual} condition. 
With two conditions and two evaluation tasks (Bias, Politeness), each condition included one of the tasks. Order of the conditions and tasks was counterbalanced, yielding four sequences.
Participants were instructed to think-aloud while completing the task using \texttt{EvalAssist}. Throughout the task, user interactions were logged. Participants were instructed to run the evaluation at least once, with the option to run it as many times as they wished. Among the models that were internally available to \texttt{EvalAssist}, we chose to use Llama 3.3-70b\footnote{\url{https://huggingface.co/meta-llama/Llama-3.3-70B-Instruct}} because our preliminary tests revealed that it performed reasonably well and fast enough for both synthetic test data generation and evaluation. 

Following each condition, participants completed a post-task survey, which guided a semi-structured interview on their perceptions of data diversity, quality, relevance, satisfaction, and helpfulness; their satisfaction with criteria and refinement strategies; and their perceived task load.
The session concluded with a final semi-structured interview, prompted by an exit survey, to explore their overall preferences comparing the tool and the manual approach. The key questions are listed in Table~\ref{tab:questions}. 
Participants received a reward equivalent to USD 37.5 upon completion.

\subsection{\texttt{EvalAssist} Setup}\label{sec:evalassist_setup}
\texttt{EvalAssist} supports two modes: 1) direct assessment, where the LLM judge selects a value or option from a rubric, and 2) pairwise comparison, where it chooses a preferred response from a data pair. We used the direct assessment mode for the user study, because people preferred this mode more than the pairwise comparison mode~\cite{ashktorab2024aligning}.
As listed in Table~\ref{tab:domain_personas}, we provided six domain examples (e.g., News Media, Healthcare, Finance, Online Knowledge Sharing, Customer Service, Acadmic Discussion), each with five personas (e.g., News Media domain: Objective Reporter, Opinion Columnist, Partisan Journalist, Sensationalist Reporter, Propagandist). We generated these domains and personas using GPT-4.1 Mini to exhibit varying levels of bias or politeness. The list of domains, personas, and prompts we used are listed in Appendix~\ref{appendix:prompt_domain_persona}.
For the length, we defined short as 1-2 sentences, medium as 3-5 sentences, and long as 5-9 sentences, which were reasonable data lengths to review within the study time.
We anticipate a more flexible implementation in future versions of the tool, allowing users to freely define their domain, persona, and length in natural language or providing more options.

\subsection{Data Analysis
}\label{sec:analysis}

We analyzed the qualitative data from interviews and open-ended survey questions using thematic analysis~\cite{BraunClarke2006}. Two researchers first independently coded the entire dataset to generate initial codes and themes. Following this, they discussed, refined, and consolidated these themes together, resolving all discrepancies until full consensus was achieved.  To calculate the number of participants who mentioned the themes, one researcher assigned the final themes to the data, and the second researcher reviewed the assigned themes. The two resolved all disagreements before calculating the final counts.

We analyzed the Likert-scale ratings  from our post-task survey using cumulative link mixed models (CLMM), a method suited for ordered data without assuming normality, implemented via the \texttt{clmm} function in the \texttt{ordinal} R package~\cite{R-ordinal}. Note that our statistical analysis using survey responses is intended to corroborate and supplement the primary qualitative findings, rather than to establish generalizable claims. 
For our primary analysis, we compared the two main experimental conditions using a two-level independent variable, Condition: \textit{Synthetic} and \textit{Manual}.
To analyze the effect of Condition on the ordinal dependent variables, we constructed a CLMM with a covariate for the task type (Bias, Politeness) and a random intercept for the participant ID. A likelihood ratio test (LRT) was conducted to calculate statistical significance (p-value), with values less than 0.05 considered statistically significant.
Effect sizes are reported as odds ratios (OR) with 95\% confidence intervals (CI)~\cite{szumilas2010explaining}. 

If the primary analysis revealed no statistical significance, we performed an exploratory analysis to determine whether the method of test case creation within the \textit{Manual} condition influenced the outcome. 
Among 24 participants, 13 participants chose to use examples we provided during the \textit{Manual} condition (\textit{Sourced}), compared to the 11 participants who wrote test cases from scratch (\textit{Authored}).
Hence, we tested a three-level independent variable, Subgroup: \textit{Synthetic}, \textit{Authored} (those who wrote data from scratch), \textit{Sourced} (those who selected data from the given real-world data). Similar to the primary analysis, we constructed a CLMM with the same covariate and random effect, conducted an LRT, and reported effect sizes in the same manner. 
When the LRT indicated a significant effect, we conducted post-hoc tests with a Bonferroni adjustment using the \texttt{emmeans} R package~\cite{R-emmeans} to identify which specific subgroups differed. 

We analyzed user behavior data, using one of two statistical tests based on whether the assumptions of normality, linearity, and homogeneity of variance were met. When the assumptions were met (e.g., alignment score), we fitted a linear mixed-effects model (LMM) using \texttt{lme4} R package~\cite{bates2025lmer} with the task type as a covariate and the participant ID as a random intercept and used an LRT to assess statistical significance. We calculated Cohen's $f^2$ to measure the effect size. If the assumptions were not met (e.g., number of test cases), we used the Friedman rank sum test, which is a non-parametric method to compare between within-subject conditions. We calculated Kendall's $W$ to measure the effect sizes. Effect sizes are interpreted based on~\cite{cohen2013statistical}. If we don't find significant difference, we compared two subgroups in the Manual condition as an exploratory analysis, using the Kruskal-Wallis rank sum test and $\eta^2_H$ for effect size.
$\eta^2_H$

\section{Results}\label{sec:results}
In this section, we answer our research questions regarding preference of the tool, synthetic data, task load, and criteria refinement, using both qualitative and quantitative analyses.
Descriptive statistics, likelihood ratio tests, and effect sizes for survey responses are reported in Table~\ref{tab:stats}. Here, we report the primary findings of interest.

\subsection{Preference (RQ1)}\label{sec:results_preference} 

\begin{table}[h]
\small	
\centering
\caption{Themes underlying participants' reasons for preferring the synthetic data generation tool. }
\label{tab:preference}
\resizebox{\textwidth}{!}{
\begin{tabular}{ll}
\hline
\toprule
\textbf{Theme} & \textbf{Description} \\
\midrule
Fast generation (N=8) & Reduced the time required to create test data manually. \\
Scalability (N=3) & Enabled users to evaluate a larger number of test data than would be feasible manually.\\
Diversity (N=7) & Provided a broader range of test data across topics, tones, vocabulary, and sentence structure. \\
Different options (N=4) & Able to generate multiple test data with varying expected outcomes, particularly borderline cases.\\
Reduce task load (N=5) & Eliminating or reducing the cognitive burden of creating test data or the manual labor of sourcing test data.\\
Customization (N=6) & Easy to iterate and edit the test data to align with users' preferences \\
Inspiration (N=3) & Introduced new ideas and perspectives that they might not have considered. \\
High-quality text (N=2) & Produced high-quality test data without grammar errors.\\
Easy to use (N=2) & Intuitive user interface and a clear presentation of information. \\
\bottomrule
\end{tabular}
}
\end{table}

The majority of participants (83\%) expressed a clear preference for generating test cases using the synthetic data tool in the \textit{Synthetic} condition ($\chi^2(2)=27, p<.05$). Nine themes emerged to explain their preferences, as listed in Table~\ref{tab:preference}.
Participants highlighted that the tool's \textbf{fast generation} and \textbf{scalability}, with one participant stating it ``\textit{allows me to produce much more test data in a shorter amount of time}'' [P24]. Another key factor was the  \textbf{diversity} of test cases. Participants noted the tool could ``\textit{generate a more diverse set of test cases than a user would by creating them manually}'' [P19], including diverse ``\textit{lexical and syntactic structures}'' [P18], ``\textit{domains and personas}'' [P23],  and ``\textit{tones}'' [P6]. 
Furthermore, participants valued the ability to generate data with \textbf{different expected outcomes} including the \textbf{borderline}  option: ``\textit{You can generate different quantities of responses with different expected criteria}'' [P22], ``\textit{especially in the case of borderline test cases}'' [P19].
Participants also highlighted the \textbf{customization} enabled by the AI's direct manipulation toolbar. For instance, one participant said, ``\textit{I like highlighting the words to select `Elaborate' after the responses are created}'' [P22].
P1 provided a clear summary of how they used the tool to diversify domains and options, customize data length, and identify blind spots: \begin{quote}``\textit{I tried a variety of domains to see how the model would respond to different domains. I kept the length short-medium because the long generations were much longer than expected. I tried to generate a fair number of examples, especially of neutral or borderline cases to see where the model distinguished between polite and neutral}'' \end{quote}
\textbf{Reducing the task load} was another common reason for their preference. P21 said ``\textit{I didn't have to manually come up with or search for test data before I can test my criteria}.'' Others said ``\textit{generating synthetic test case helps us to have the test cases prepared automatically} [P8],'' ``without thinking of the grammar, structure etc.'' [P9].


In contrast, two participants favored the manual creation or selection of test cases in the \textit{Manual} condition. They raised concerns about the tool's limitations, suggesting its outputs may lack diversity in large-scale generations, and are only helpful when a task fits neatly into a predefined domain. A few participants also found value in using real-world examples. P11 preferred creating their own data because they ``could come up (with examples) that the tool might not have considered,'' while P12 preferred selecting from the provided real-world data to ``see how accurately the (EvalAssist) tool would react to real-life examples''.

The other two participants had no preference, viewing the two methods, \textit{Synthetic} and \textit{Manual}, as complementary; they suggested that manually creating test cases could yield useful real-life examples, while automated generation remains valuable for its speed and scalability.

When asked which specific features in the synthetic data generation tool were most helpful, the \textbf{Synthetic instance generation feature} (N=17) was overwhelmingly cited as the most helpful feature (Section~\ref{sec:direct_ai_manipulation}). Participants particularly valued the ability to specify the domain (N=10), persona (N=7) and quantity (N=6), particularly the borderline cases (N=4). P3 noted, ``\textit{it's easy to forget some of these, if I were to produce personas and domains myself.}" 
The second helpful feature was the \textbf{Direct AI manipulation toolbar} (N=8), within which the Elaborate function (N=3) was the most cited. P22 articulated ``\textit{Elaborate was the most helpful feature of the tool because it took parts of the sentence that I thought were confusing}." This indicates that users sometimes need assistance understanding data that they did not create themselves.

\subsubsection{Summary}
Participants preferred the synthetic data generation tool to manual approaches, 
mainly due to its speed, data diversity, and customization options. Of all the features, the ability to generate synthetic instances by configuring their domain, persona, and quantity was found to be the most helpful.

\subsection{Data (RQ2)}\label{sec:results_data} 
\begin{table*}[t]
\centering
\small

\caption{Comparison of survey outcomes for the primary (\textit{Synthetic} vs. \textit{Manual}) and exploratory (\textit{Synthetic} vs. \textit{Authored} vs. \textit{Sourced}) analyses. Descriptive statistics for ordinal values (medians (M) with interquartile ranges (IQR) in parentheses), and odds ratios (OR) with 95\% confidence intervals with the \textit{Synthetic} condition as the reference are reported. Likelihood ratio test (LRT) results ($\chi^2$ and $p$-value) are also reported, showing that only the syntactic diversity differed significantly across the subgroups (bolded).}
\label{tab:stats}
\resizebox{\textwidth}{!}{
\begin{tabular}{l lllllllll}
\toprule
 &  & \multicolumn{3}{c}{\textbf{Condition (2-level)}} & 
\multicolumn{5}{c}{\textbf{Subgroup (3-level)}} \\
\cmidrule(lr){3-5} \cmidrule(lr){6-10}
\textbf{RQ} &  \textbf{Synthetic} & \multicolumn{2}{l}{\textbf{Manual}}  & \textbf{LRT} & \multicolumn{2}{l}{\textbf{Authored}}  & \multicolumn{2}{l}{\textbf{Sourced}}  &  \textbf{LRT} \\
Question & M (IQR) & M (IQR) & OR [95\% CI] & $\chi^2 (p)$ & M (IQR) &  OR [95\% CI] & M (IQR) & OR [95\% CI] &  $\chi^2 (p)$ \\

\midrule
\multicolumn{10}{l}{\textbf{Data}} \\ 
Topical Diversity     & 4 (4--5) & 4 (3--4.25) & 0.42 [0.14, 1.28] &  2.44 (.19) & 4 (3--4)  & 0.30 [0.08, 1.15] &  4 (4--5)& 0.63 [0.16, 2.54] & 3.24 (.19) \\
Lexical Diversity     & 4 (3.75--5) & 4 (3--5)    & 0.56 [0.18, 1.73] & 1.04 (.31)  & 3 (2--4.5) & 0.26 [0.06, 1.21]  & 4 (4--5) & 1.04 [0.27, 4.02] & 3.07 (.22) \\
Syntactic Diversity   & 4 (4--5) & 4 (3.75--5) & 0.56 [0.16, 1.91] & 0.89 (.35)  & 3 (2.5--4)  & 0.12 [0.02, 0.68]  & 4 (4--5)  & 2.48 [0.50, 12.33] & \textbf{7.92 (<.05)}\\
Option Diversity   & 4.5 (4--5)  & 4 (4--5)   & 1.09 [0.73, 1.62] & 0.03 (.87)  & 4 (4--5) & 0.74 [0.50, 1.09]  & 5 (4--5) & 1.54 [1.04, 2.27] & 0.84 (.66) \\
Helpfulness           & 5 (4--5) & 5 (4--5)    & 1.73 [0.44, 6.70] & 0.64 (.43)  & 4 (4--5) & 0.54 [0.12, 2.43] & 5 (5--5)  & 8.30 [0.87, 79.45] & 5.30 (.07) \\
Satisfaction          & 4.5 (4--5) & 5 (4--5)    & 0.90 [0.29, 2.76] & 0.04 (.85)  & 4 (3.5--5) &   0.64 [0.16, 2.61]  & 5 (4--5) & 1.26 [0.30, 5.27] & 0.63 (.73) \\
Quality               & 4 (3--4) & 4 (4--5)    & 1.84 [0.60, 5.62] & 1.17 (.28)  & 4 (4--4) & 1.13 [0.28, 4.48]  & 4 (4--5) & 3.11 [0.74, 13.13] & 2.61 (.27) \\
Relevance             & 5 (4--5) & 5 (4--5)    & 0.80 [0.21, 3.02] & 0.11 (.74)  & 5 (4--5) & 0.83 [0.15, 4.52]  &   5 (4--5) & 0.76 [0.15, 4.00] & 0.12 (.94) \\
\midrule
\textbf{Task Load} & & & & & & \\
Mental Demand        & 3 (3--3) & 3 (2.75--3)    & 0.31 [0.08, 1.24] & 3.04 (.08)  & 3 (2.5--3) & 0.24 [0.04, 1.55] & 3 (3--3) & 0.39 [0.07, 2.12] & 3.23 (.20) \\
Performance          & 4 (3--4) & 4 (3--4)    & 0.99 [0.32, 3.13] & 0.00 (.99)  & 4 (3--4) & 0.70 [0.15, 3.26]  & 4 (4--4) & 1.33 [0.32, 5.61] & 0.45 (.80) \\
Effort               & 3 (2--3) & 3 (2--3)   & 0.18 [0.12, 0.26] & 1.53 (.22) & 3 (2.5--3) & 1.31 [0.01, 130.41]  & 3 (2--3) & 0.88 [0.00, 3.39] & 3.57 (.17)  \\
\midrule
\textbf{Criteria} & & & & & & \\
Satisfaction         & 4 (4--5) & 5 (4--5)    & 2.01 [0.59, 6.83] & 1.30 (.25)  & 5 (4--5) &  2.51 [0.45, 14.05]  & 5 (4--5) & 1.67 [0.35, 7.97] & 1.44 (.49) \\
\bottomrule
\end{tabular}
}
\end{table*}

\subsubsection{Quantity}
We calculated the \textbf{number of test cases} used for evaluation. When using the synthetic data generation tool, participants generated significantly more test cases than they did with the manual approach (Synthetic: M=9.08, SD=4.99; Manual: M=5.21, SD=2.04). A Friedman rank sum test revealed a statistically significant difference in the number of test cases between \textit{Synthetic} and \textit{Manual} conditions ($\chi^2$(1) = 14.73, $p < .001$), with a large effect size (Kendall's $W$ = 0.61, 95\% CI = [0.27, 0.96]). 

We also calculated the total\textbf{ number of words per test case}. 
We found that test cases generated in the \textit{Synthetic} condition  (M=32.21, SD=13.23) generated significantly more words compared to the \textit{Manual} condition (M=16.24, SD=4.62). A Friedman rank sum test revealed a statistically significant difference in the number of words between \textit{Synthetic} and \textit{Manual} conditions ($\chi^2$(1) = 13.5, $p<.001$), with a large effect size (Kendall's $W$ = 0.56, 95\% CI = [0.18, 1]).  
As a supplementary analysis, we analyzed task completion time and found no meaningful differences between conditions ($\chi^2$(1) = .52, $p=.47$; $f^2$ = 0.10). This may suggest that more test data were generated in the \textit{Synthetic} condition than the \textit{Manual} condition within a comparable amount of time.

\subsubsection{Syntactic Diversity}
While our survey responses on perceived syntactic diversity did not show a statistically significant difference between the \textit{Synthetic} and \textit{Manual} conditions, the exploratory analysis revealed a significant difference in syntactic diversity among the subgroups (see Table~\ref{tab:stats}). Post-hoc tests showed that test cases produced in the \textit{Synthetic} subgroup were perceived to be more syntactically diverse than those in the \textit{Authored} subgroup ($p<.05$). Likewise, test cases in the \textit{Sourced} subgroup were perceived to be significantly more syntactically diverse than those in the \textit{Authored} subgroup ($p<.05$).  We found no significant difference in the perceived syntactic diversity between the \textit{Synthetic} and \textit{Sourced} subgroups ($p=.51$). This result suggests that test cases generated by our tool are perceived to be as syntactically diverse as real-world data, and more syntactically diverse than hand-crafted data.

\subsubsection{Topical Diversity}
We examined how participants used the Synthetic instance generation feature in the \textit{Synthetic} condition. Participants selected a limited number of parameters, using 1.50 domains (SD=0.88), 1.75 personas (SD=0.99), and 1.21 length options (SD=0.51) on average. 
While there were a few participants who explored multiple parameters to seek for diversity, many participants deliberately selected a challenging domain and persona in order to test their criteria. For instance, after conducting the bias evaluation task, P6 said: ``\textit{I chose Health (domain) since it's a topic where bias is especially hidden. I chose specifically a healthcare conspiracy theorist (persona) since it could seem like an easy choice to achieve a `biased' text, but I wanted to test the criteria to actually achieve `non-biased' text.}'' Other common reasons underlying their parameter choices were familiarity and time efficiency. For example, P16 articulated the rationale behind their choices: ``\textit{familiarity with the domain, the most dominant persona in the domain, and wanted to keep it short for better (quicker) evaluation}.''   
This behavior suggests that participants found these parameters useful for customization, allowing participants to tailor domains, personas, and data lengths based on their preferences. While a few participants explored multiple parameters for diversity, some didn't because they focused on testing the criteria within a few carefully chosen domains and personas and prioritized efficiency. This finding may explain why the survey responses indicated that the \textit{Synthetic} and \textit{Manual} conditions had similar perceived topical diversity.

\subsubsection{Lexical Diversity}
We calculated both the \textbf{number of unique terms} and the \textbf{ratio} (the average number of unique terms used, divided by the total number of words) in each test case to capture lexical diversity while accounting for potential length effects in the generated data.
We found our synthetic data tool (M = 32.21, SD = 13.23) generated significantly more unique terms compared to the manual approach (M = 16.24, SD = 4.62).  A Friedman rank sum test revealed a statistically significant difference between the conditions ($\chi^2$(1) = 13.5, $p$ < .001) with a large effect size (Kendall's $W$ = 0.56, 95\% CI = [0.25, 1]). 
However, when considering the data length, we found that test cases generated in \textit{Synthetic} condition (M=0.83, SD=0.07) generated a significantly lower ratio of unique terms, compared to the \textit{Manual} condition (M=0.88, SD=0.05). A Friedman rank sum test revealed a statistically significant difference between the conditions ($\chi^2$(1) = 8.17, $p$ < .001) with a moderate effect size (Kendall's $W$ = 0.34, 95\% CI = [0.06, 0.84]). 
Conflicting results between the absolute count and the ratio indicate that while the tool generated more unique terms overall, the relative diversity decreased due to increased text length. This may account for why the \textit{Synthetic} condition had comparable perceived lexical diversity to the \textit{Manual} condition, according to the survey. 

\subsubsection{Option Diversity}
Unlike other parameters, participants made more extensive use of the ``quantity per option'' parameter in the configuration tool, covering an average of 3.25 options (SD = 1.02) out of the four available options (polite, neutral, impolite, borderline), and 2.92 options (SD = 0.29) out of three options (biased, non-biased, borderline). 
Many participants also mentioned that they aimed for option diversity, as P24 said ``\textit{try to generate a diverse test data, where there's at least one example of each: polite, impolite, neutral, or borderline.}'' Some participants focused on generating challenging options ``\textit{I tried to generate a fair number of examples, especially of neutral or borderline cases to see where the model distinguished between polite and neutral}'' [P1]. 
Similarly, in the \textit{Manual} condition, many participants in the \textit{Authored} subgroup mentioned that they tried to write test cases with varied expected outcomes: ``\textit{When creating test data, I thought first to modify the existing test data in a way so that the evaluator would provide the opposite response.}'' [P24]. 
In contrast, participants in the \textit{Sourced} subgroup often ``\textit{randomly picked a test sentence}'' [P5] from the ten examples we provided. This strategy could potentially reduce the option diversity in real-world contexts when practitioners use datasets with unbalanced class distribution. However, in this experiment, we provided examples with varying predicted outcomes, which may have led to the lack of a significant difference in perceived option diversity across conditions.

\subsubsection{Quality, Relevance, Helpfulness, and Satisfaction}
As shown in Table~\ref{tab:stats}, the median ratings for perceived quality, relevance, helpfulness, and satisfaction were consistently high (4 or higher on a 5-point Likert scale) across all conditions and subgroups, with no statistically significant differences found. These results suggest that the proposed tool generates test data that is as effective as hand-crafted or real-world data regarding these dimensions.

\subsubsection{Summary}
The synthetic data generation tool successfully enhanced data quantity as it generated significantly more and longer test cases than they did with manual methods. 
While the tool improved syntactic diversity over hand-crafted data, other diversity dimensions (e.g., lexical, topical, option) were not significantly different between conditions. 
The tool maintained high levels of perceived data quality, relevance, helpfulness, and user satisfaction, similar to those from the manual approach.

\subsection{Task load (RQ3)} 

From our qualitative findings in Section~\ref{sec:results_preference}, we found that participants favored the tool because it reduced their workload in writing or selecting test cases. 
However, as shown in Table~\ref{tab:stats}, the median ratings for task load (mental load, performance, and effort) were moderate (3 or 4 on a 5-point Likert scale) across all conditions and subgroups, with no statistically significant differences found. These results suggest that using the tool did not significantly increase or decrease users' perceived workload.
As reported in Section~\ref{sec:results_data}, significantly more test cases were used in the \textit{Synthetic} condition compared to the \textit{Manual} condition, and the synthetic data were longer on average, requiring more effort to review and interpret. Thus, the tool may have reduced per-case workload, while the total effort across all cases was comparable to that of the manual approach.

\subsubsection{Summary}
Perceived task load was not significantly different between the \textit{Synthetic} and \textit{Manual} conditions, 
likely because the reduced effort per test case was offset by participants creating a much larger volume of longer data using the tool.

\subsection{Criteria (RQ4)} 
\subsubsection{Satisfaction}
Participants were highly satisfied with the final criteria they saved across all conditions, as demonstrated by high median scores (above 4 on a 5-point Likert scale) in Table~\ref{tab:stats}. No statistical difference was found across conditions.

\subsubsection{Criteria refinement strategy}
By analyzing system logs, interviews, and survey responses, we identified eight main strategies participants used to refine their evaluation criteria (Table~\ref{tab:criteria_strategy}). Overall, many strategies were shared between the conditions, which suggests that the test data had a limited downstream impact on the criteria refinement process. The most common strategy across all conditions was the \textbf{Defining concepts} strategy, where they defined their criteria and options using related synonyms and concepts (e.g., factuality, fairness). For example, P10 described that they changed the Bias option description to be ``\textit{not strictly factual and expressing an opinion.}''

In the \textit{Manual} condition, participants more frequently used the \textbf{Grounding in specific data or domain} strategy. For instance, P5 copied and pasted a specific test instance into the Non-biased description. Additionally, more participants in \textit{Manual} condition used the \textbf{Rule-based requirement} strategy. For example, P1 imposed a single-sentence rule to the Impolite option based on one or a few examples they had reviewed: ``\textit{Non-biased: Is it just one sentence lacking additional context that might be helpful or relevant?}" These strategies can result in overly specific criteria that are not generalizable to other data. 
They occurred less frequently in the \textit{Synthetic} condition, potentially because the variety of synthetic data exposed participants to a broader range of examples, likely discouraging over-reliance on a single example. 

\begin{table}[h]
\small	
\centering
\caption{Strategies participants used to refine the criteria, followed by the number of participants from the \textit{Synthetic} and \textit{Manual} conditions who used each strategy. Overall, participants used similar strategies in both conditions. The ``Grounding in specific data or domain'' and ``Rule-based requirements'' strategies were used by  more participants in the \textit{Manual} condition (bolded) than the \textit{Synthetic} condition.}
\label{tab:criteria_strategy}
\begin{tabular}{p{0.2\textwidth}p{0.6\textwidth} cc}
\hline
\toprule
\textbf{Theme} & \textbf{Description} & \textbf{Synthetic} & \textbf{Manual} \\
\midrule
Defining concepts & Clarifying the meaning of a criterion or its options, often by providing synonyms or related concepts (e.g., fairness or factuality to describe the Bias criterion) &14&15\\
Tone and language& Defining the desired tone or language style for each criterion option (e.g., sarcastic tone, respectful language). &9&8\\
Grounding in specific data or domain	& Copy and pasting a specific example from the test data to describe a criterion option, or focus on a particular domain (e.g., politics).&1&\textbf{6}\\
Rule-based requirements	& Defining objective, often quantitative or programmable rules, such as formatting or length constraints.&2&\textbf{6}\\
Clarifying nuanced cases&Refining the criteria to better handle ambiguous, borderline, or complex examples. &3&2\\
Specific keywords & Incorporating exemplary words or phrases (e.g., use words like ``please'' to indicate politeness).&5&6\\
Using explanations & Leveraging the explanations under generated results to improve the criteria. &2&0\\
No change & Leaving the criteria unchanged. &5&4\\
\bottomrule
\end{tabular}
\end{table}

\subsubsection{Evaluation Alignment}

To assess how well participants' final criteria aligned with their expectations, we calculated the number of agreements between the participant's expected result and the \texttt{EvalAssist}'s generated result during the task. Then we divided the number to the total number of test cases to calculate the \textbf{agreement score}. The agreement scores were similar between the \textit{Synthetic} condition (M = 0.85, SD = 0.23) and the \textit{Manual} condition (M = 0.89, SD = 0.2). A Friedman rank sum test revealed no statistically significant differences between the conditions ($\chi^2(1)$
= 0.6, $p = .44$; Kendall's $W$ = 0.025, 95\% CI = [0, 0.27]), and between the \textit{Manual} condition's subgroups ($\chi^2(1)$ = 0.04, $p = .84$; $\eta^2_H$ = -0.04, 95\% CI = [-0.05, 0.18]).

We also measured the probability that  \texttt{EvalAssist}'s generated result using the user's final criteria would match the participant's own manual evaluation results on the validation data (i.e., \textbf{alignment score}). The \textit{Synthetic} condition's average alignment score was 0.57 (SD=0.15), while  \textit{Manual} condition's average alignment score was 0.59 (SD=0.13). No significant difference was found in this alignment score between the two experimental conditions ($\chi^2(1)$ = 0.46, $p = .50$; $f^2$ = 0.14), 
and between the \textit{Manual} condition's subgroups ($\chi^2(2)$ = 2.11, $p = .35$; $f^2$ = 0.25). 


\subsubsection{Summary}
Both the synthetic data generation and manual approaches resulted in comparable levels of user satisfaction with the criteria and its alignment with participant expectations. 

\section{Discussion}\label{sec:discussion}
In this section, we discuss the utility of the synthetic data generation tool as well as opportunities for improvement. We also discuss designs to better support the criteria refinement and alignment process.

\subsection{Synthetic Data Generation in LLM-as-a-Judge Workflow}
Ensuring the quality and diversity of synthetic data is an active research area, and our work has demonstrated the effectiveness of the human-in-the-loop method where users either edit prompts used for generation or post-process the generated data. 
The \textbf{synthetic instance generation} feature was the most helpful and frequently used by participants, as it gave participants direct control over the generation prompts. This finding supports prior research that identifies controllability as a critical factor in the user acceptance of AI systems~\cite{roy2019automation,kocielnik2019will}. 
Some participants also found the \textbf{direct AI manipulation} toolbar helpful. This feature aligns with Variation Theory~\cite{runesson2006possible}, which posits that students learn through experience varying critical aspects. 
\citet{gebreegziabher2025supporting} also found that counterfactual variations are useful for the user's reflection. 
Our feature embodies this theory and extends this prior work by allowing users to choose which part of the text to vary and how (e.g., paraphrase, elaborate), thus providing greater flexibility and learning opportunity.

Participants perceived the tool as \textbf{a faster and more scalable way} to generate test data. This perception was corroborated by behavioral logs, which showed that participants generated significantly more and longer test cases when using the synthetic data generation tool compared to the manual approach. Even with a large volume of test data users need to generate and review, this increase did not appear to raise the \textbf{perceived workload}. 
Moreover, participants found the synthetic data to have comparable \textbf{quality, relevance, and helpfulness and satisfaction} to hand-created or real-world examples.

Moreover, participants valued the tool for \textbf{customization, diversity, and inspiration}. They appreciated the ability to specify or explore multiple parameters like domains, personas, and data length, and to further edit the generated data with AI assistance. Participants reported that this process not only helped them tailor data to their needs, but also \textbf{sparked new ideas} and helped \textbf{identify their blind spots}, particularly through the generation of borderline cases. 

Our study showed that the generated synthetic data was as effective as hand-crafted or real-world data for helping participants refine their evaluation criteria until satisfied, with their \textbf{criteria satisfaction} remaining consistently high across all conditions. This high satisfaction may be attributed to the ``IKEA effect," an increase in valuation of self-made products~\cite{norton2012ikea}. Regardless of how the data was generated, participants still had to refine the criteria themselves, which may have triggered the effect and led to high satisfaction.

\subsection{Opportunities for Improving Data Diversity}
Despite its benefits, we also identified clear opportunities for improvement in synthetic data generation. When a user generates multiple data under the same configuration, the tool sometimes produced mostly repetitive outputs despite high-temperature settings, a known issue with current LLMs when using similar prompts~\cite{yeh2025exploring}.  Additionally, the generated text lacked the ``messiness'' of real-world data. LLMs tend to produce well-structured, grammatically correct sentences, whereas human data often contains slang, jargon, informal language, and grammar errors. Addressing these gaps to increase the lexical and syntactic diversity of the generated synthetic data is a crucial direction for improving the tool, such as using multiple LLMs with different fine-tuning methods~\cite{sun-etal-2024-toward}. 

The tool enabled participants to generate synthetic data that covered a diverse set of outcomes for their evaluation criteria. In contrast, the \textit{Sourced} subgroup typically chose instances randomly from a provided dataset. Hence, we anticipate that our tool could be a useful solution for addressing real-world data challenges, such as having imbalanced class distribution. However, we found no significant difference in the option diversity across conditions. This is likely a limitation of our task design, as each evaluation task included only two or three options, restricting the potential for large differences to emerge. Future studies should explore different evaluation tasks with more granular options or scores to better assess how the tool supports generating diverse options. 

Furthermore, we anticipate that diversity of synthetic test data could be further improved through effective choice architecture, which explains how choices are presented influences decisions~\cite{thaler2021nudge}. For example, participants selected multiple options, likely because the interface listed all available options and defaulted the number of data instances to 1 per available option. In contrast, other parameters (e.g., domain, persona, and text length) were implemented as dropdowns that restricted participants to a single choice per generation. Implementing similar interface design for these parameters that allows participants to select multiple domains, personas, and lengths at a time, could encourage participants to explore a wider range of options.

\subsection{Criteria Refinement and Sense-making}

While participants generated significantly more test cases with the tool, this did not necessarily lead to a better criteria compared to the \textit{Manual} condition. 
Also, the final alignment between human and LLM judges using their criteria were comparable across conditions.
These findings can be interpreted through the lens of Sense-making Theory~\cite{weick1995sensemaking}, which models how individuals search for information, identify a knowledge gap, and synthesize information to support decisions or actions~\cite{zhang2015conceptual}. Our tool successfully supported the initial information seeking stage of this process by providing users with more data. However, more data also required greater analytical ability of users to identify misalignment patterns and translate them into actionable strategies for criteria refinement. The lack of support in our tool for these subsequent stages led participants to adopt divergent refinement strategies. 

To bridge this gap, future design of the LLM-as-a-judge system 
should integrate methods that help users to gain insights across generated data, and suggest appropriate refinement strategies. We could draw inspirations from prior work~\cite{leeevaligner,gebreegziabher2025metricmate,kim2024evallm}, including 
\texttt{EvalGen}~\cite{shankar2024validates} that suggests an initial set of criteria for users to further edit. By incorporating such guidance, future systems can more effectively support users' sense-making process and lead to greater impact on human-LLM alignment.

\subsection{Limitations and Future Work}
Our study recruited participants from a single company. While we were able to recruit individuals with relevant experience or interest in data evaluation in this population, 
future work should examine a broader population to improve generalizability.
We focused on a text generation task using the \texttt{EvalAssist} framework with subjective criteria and textual data. Future studies could extend this work by examining a broader range of evaluation tasks (e.g., summarization), framework (e.g., multi-modal evaluation framework), criteria (e.g., objective), and types of data (e.g., images).
Lastly, the survey statistical results should be interpreted alongside other results and are not intended to make standalone claims due to relatively small sample size.
In our exploratory analysis, participants self-selected into the \textit{Authored} and \textit{Sourced} subgroups. Therefore, any differences observed between these two subgroups could be influenced by their pre-existing characteristics (e.g., proficiency in writing).

\section{Conclusion}\label{sec:conclusion}
Users are often constrained by the lack of diverse, high-quality test data that would help them create subjective criteria in an LLM-as-a-judge evaluation task. 
Based on a formative study, we designed and developed a synthetic data generation tool for an LLM-as-a-judge system, \texttt{EvalAssist}, that allows users to easily control the prompts behind the generation and post-process the generated data. 
We conducted a mixed-method user study (N=24) to test the user experience and effectiveness of the tool. Synthesizing qualitative interview data, quantitative survey ratings, and behavioral logs, 
we found that most participants preferred the synthetic data generation tool over the manual approach. Using the tool, they were able to generate significantly more, longer, and syntactically diverse test cases without sacrificing data quality or incurring additional task load. The downstream impacts of the generated data on the evaluation criteria and their alignment with users' expectations were as effective as human-crafted or real-world data. 
A promising future direction is to enhance the tool to support greater data diversity and suggest effective refinement strategies for sense-making. 

\bibliographystyle{ACM-Reference-Format}
\bibliography{reference.bib}


\newpage
\appendix

\section{\texttt{EvalAssist} User Interface}\label{appendix:UI}
\begin{figure}[H]
\centering
  \includegraphics[width=\textwidth]{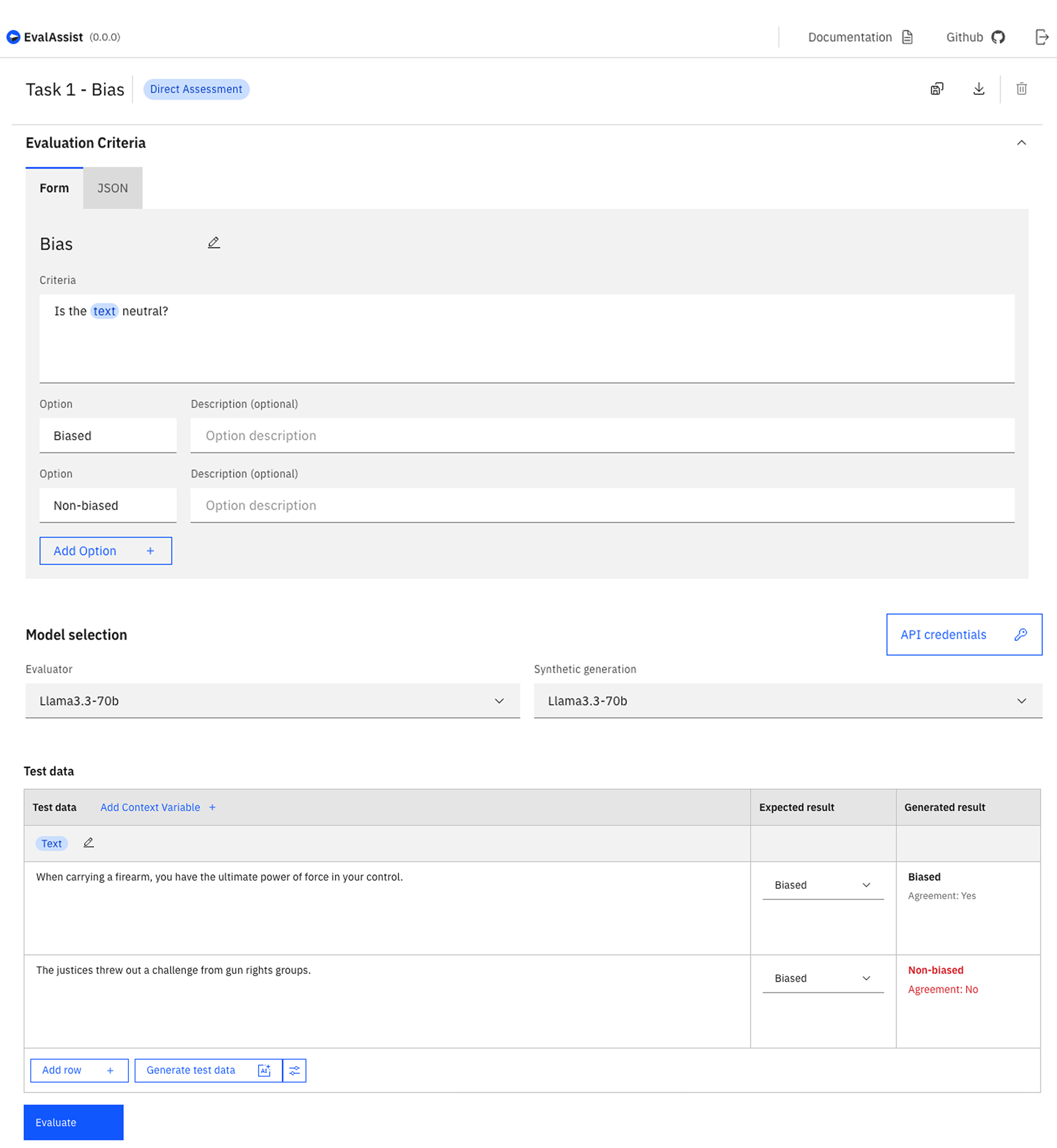}
  \caption{Screenshot of the full \texttt{EvalAssist} interface for Bias evaluation task}
  \Description{This image displays our EvalAssist user interface full screenshot used in the study. In the top section, a user defines the Evaluation Criteria, setting up a "Bias" criteria with "Biased" and "Non-biased" options as the possible outcomes. Below, they select the AI models for both evaluation and synthetic data generation. The bottom table displays the Test data, where the user provides example sentences and their expected result, and the system shows the generated result, highlighting agreement or disagreement in red between the human and the judge. There's an "Add row" button and "Generate test data" button below the test data. At the very end, there's an "Evaluate" button highlighted in blue.}
  \label{fig:evalassist}
\end{figure}

\begin{figure}[H]
\centering
  \includegraphics[width=\textwidth]{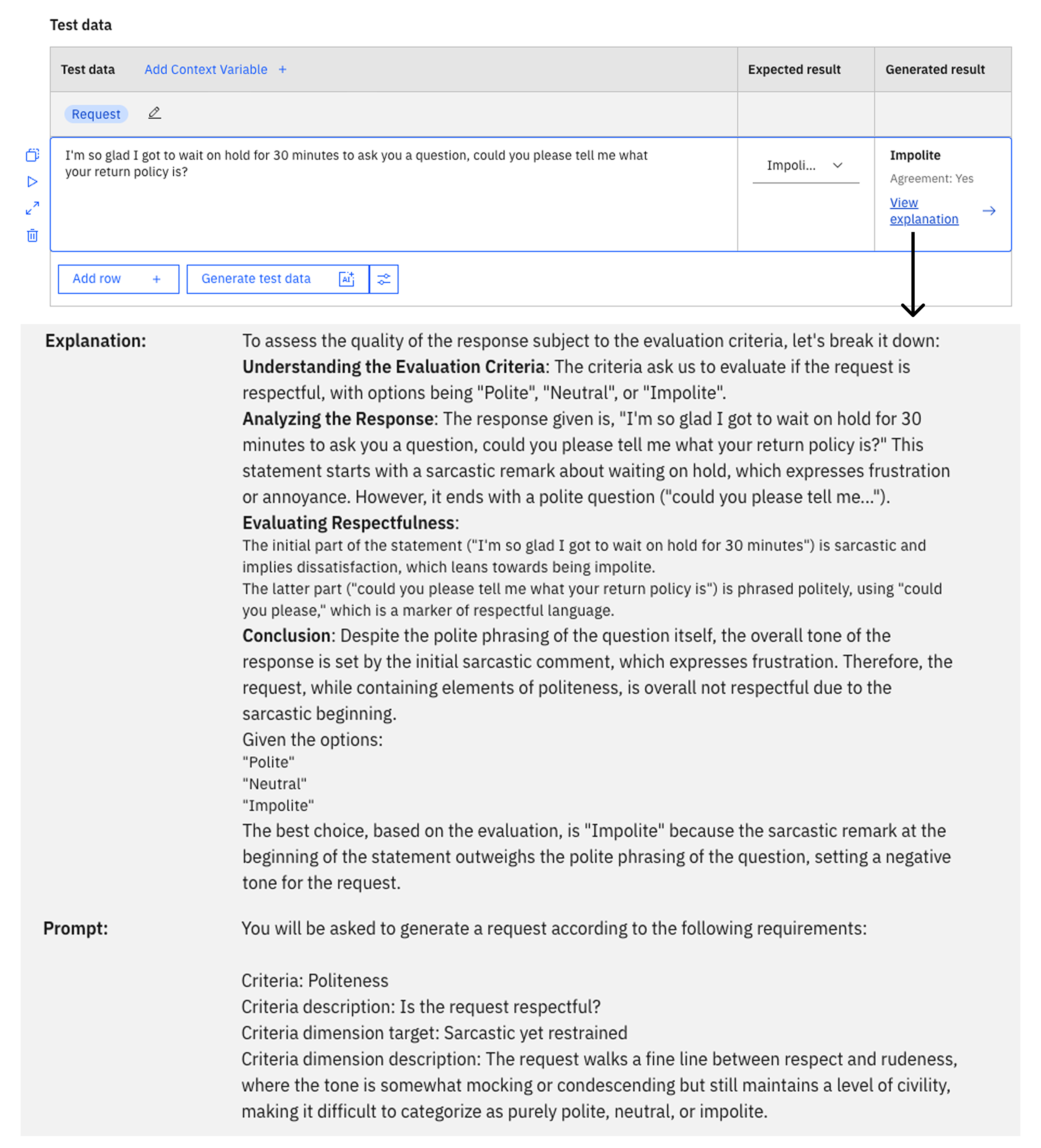}
  \caption{Explanation of a synthetic data instance . When a user hovers over a specific test instance, a ``View explanation'' button beneath the generated result will appear. Clicking the button will open a pop-up window that displays the evaluation rationale and the prompt used to generate the instance. Explanation shortened for readability.}
  \Description{This screen capture shows the EvalAssist user interface, specifically the test data section. When a user hovers over a specific test instance, several buttons appear on the left side (e.g., Duplicate, Evaluate, See Details, Remove), and a “View Explanation” button appears beneath the generated result. Clicking the “View Explanation” button opens a pop-up window that displays the evaluation rationale and the prompt used to generate the instance.}
  \label{fig:explanation}
\end{figure}

\section{Domain and Persona Generation Prompt}\label{appendix:prompt_domain_persona}
As listed in Table~\ref{tab:domain_personas}, we generated six domains, each with five personas, using outputs from the LLM (GPT-4.1 Mini) based on the prompts below. 
For consistency, the prompt included a specific domain example corresponding to the domain of the real-world examples used in the \textit{Manual} condition for consistency.

\setlength{\tabcolsep}{1mm}  
\begin{table}[h]
\small
\centering
\caption{Examples domains and personas implemented in \texttt{EvalAssist} for the user study.   }
\label{tab:domain_personas}
\resizebox{\textwidth}{!}{
\begin{tabular}{ l l }
\hline
\toprule
\textbf{Domain}  & \textbf{Persona} \\
\midrule
News Media & Objective Reporter, Opinion Columnist, Partisan Journalist, Sensationalist Reporter, Propagandist \\
Healthcare & Evidence-Based Doctor, Wellness Blogger, Alternative Medicine Advocate, Pharmaceutical Rep, Health Conspiracy Theorist \\
Finance & Certified Financial Advisor, Personal Finance Blogger, Stock Market Influencer, Cryptocurrency Enthusiast, Scam Promoter \\
Online Knowledge QA & Abrasive Commenter, Sarcastic Critic, Neutral Contributor, Encouraging Helper, Polite Moderator \\
Customer Service & Impatient Agent, Sarcastic Agent, Standard Support Rep, Friendly Support Rep, Empathetic Specialist \\
Academic Discussion & Disrespectful Debater, Passive-Aggressive Speaker, Objective Analyst, Diplomatic Scholar, Respectful Professor \\
\bottomrule
\end{tabular}
}
\end{table}

\begin{lstlisting}[breaklines=true]
What are three realistic and distinct domains, including the news media domain, along with five unique personas that could display varying levels of bias, from low to high? 
\end{lstlisting}

\begin{lstlisting}[breaklines=true]
What are three realistic and distinct domains, including the online knowledge QA domain, along with five unique personas that could display varying levels of politeness, from low to high? 
\end{lstlisting}

\section{Synthetic Data Generation Prompts}\label{appendix:prompt}

\subsection{Synthetic Instance Generation}

\textbf{Data generation for a target option:}
\begin{lstlisting}[breaklines=true]
You will be asked to generate a response according to the following requirements:

Criteria name: {CRITERIA_NAME/e.g., Bias}
Criteria description: {CRITERIA_DESCRIPTION/e.g., Is the text neutral}?
Criteria dimension target: {CRITERIA_OPTION_NAME/e.g., Biased}
Criteria dimension description: {CRITERIA_OPTION_DESCRIPTION/e.g., Considering only one perspective.}

Your task is to generate an answer that STRICTLY follows this requirement. This is for evaluation purposes.

Important:
- The generated response is going to be evaluated on the {SELECTED_DOMAIN} domain
- Adopt the following persona: {SELECTED_PERSONA}
- The generated response's length should be {SELECTED_LENGTH}
- The response should be considered to be evaluated as {CRITERIA_OPTION_NAME} based on the criteria {CRITERIA_NAME}

The output should be a markdown code snippet formatted in the following schema, including the leading and trailing "```json" and "```":

```json
{
"Response": string // the requested Response
}
```

\end{lstlisting}

\textbf{Data generation for a borderline case:}
\begin{lstlisting}[breaklines=true]
You will be provided with a criteria. The criteria is composed by a name, a description and a set of criteria options. Describe a borderline criteria option that lies between the criteria options.

Criteria name: {e.g., Bias}
Criteria description: {e.g., Is the text neutral}?
Criteria options:
    Biased: {user description e.g., Considering only one perspective.}
    Non-biased: {user description e.g., Considering multiple perspectives.}

Provide a natural language description of what it means to be a borderline case among these criteria options. Your description should mirror the style and format of the original criteria options but describe the subtle ways in which the case partially satisfies multiple criteria while not fully satisfying any single one.

The output should be a markdown code snippet formatted in the following schema, including the leading and trailing "```json" and "```":
 
```json
{
    "name": string  // the name of borderline criteria
    "description": string  // the description of borderline criteria
}
```
\end{lstlisting}
\subsection{Direct AI Manipulation Prompt}

\textbf{Rephrase:}

\begin{lstlisting}[breaklines=true]
You will be provided with:
- A selected text
- A text containing that selection, with the selection marked using <rephrase> tags
Your task is to rephrase the selected text such that:
- It preserves the original meaning and intent
- It fits seamlessly into the original text, both semantically and grammatically
- The rephrased selection must not disrupt the sentence structure or introduce grammatical errors (e.g., missing prepositions or incorrect tense).
- Do not introduce any new information that is not present in the original text.
- If the selection is equal to the whole text, your task is to rephrase the whole text.
- Examples: `toddler' changed to `kid', `terrorist' changed to `extremist', `men' changed to `human', `easy' changed to `simple', `great' changed to `excellent'

Selection:
{TEXT_SELECTION/e.g., the weather is warm and humid}

Text with selection (wrapped in-between <rephrase> tags):
{TEXT_WITH_SELECTION/e.g., On most days, <rephrase>the weather is warm and humid<rephrase>, with temperatures often soaring into the high 80s and low 90s Fahrenheit (around 31-34C).}

The output should be a markdown code snippet formatted in the following schema, including the leading and trailing "```json" and "```":
```json
{
    "response": string  // the selection to rephrase
}
Don't forget to enclose the response value in double quotes.
\end{lstlisting}

Generation result example: On most days, the weather is hot and muggy
\newline
\newline

\textbf{Elaborate:}

\begin{lstlisting}[breaklines=true]
You will be provided with:
- A selected text
- A text containing that selection, with the selection marked using <elaborate> tags
Your task is to elaborate on the selected text such that:
- It preserves the original meaning and intent
- It fits seamlessly into the original text, both semantically and grammatically
- The elaborated selection must not disrupt the sentence structure or introduce grammatical errors (e.g., missing prepositions or incorrect tense).
- Do not introduce any new information that is not present in the original text.
- If the selection is equal to the whole text, your task is to elaborate on the whole text.

Selection:
{TEXT_SELECTION/e.g., the weather is warm and humid}

Text with selection (wrapped in-between <elaborate> tags):
{TEXT_WITH_SELECTION/e.g., On most days, <rephrase>the weather is warm and humid<rephrase>, with temperatures often soaring into the high 80s and low 90s Fahrenheit (around 31-34C).}

The output should be a markdown code snippet formatted in the following schema, including the leading and trailing "```json" and "```":
```json
{
    "response": string  // the selection to elaborate on
}
Don't forget to enclose the response value in double quotes.
\end{lstlisting}

Generation result example: On most days, the weather is warm and humid, characterized by a tropical climate with high temperatures and a significant amount of moisture in the air
\newline
\newline

\textbf{Shorten:}
\begin{lstlisting}[breaklines=true]
You will be provided with:
- A selected text
- A text containing that selection, with the selection marked using <shorten> tags
Your task is to shorten the selected text such that:
- It preserves the original meaning and intent
- It fits seamlessly into the original text, both semantically and grammatically
- The shortened selection must not disrupt the sentence structure or introduce grammatical errors (e.g., missing prepositions or incorrect tense).
- Do not introduce any new information that is not present in the original text.
- If the selection is equal to the whole text, your task is to shorten the whole text.

Selection:
{TEXT_SELECTION/e.g., the weather is warm and humid}

Text with selection (wrapped in-between <shorten> tags):
{TEXT_WITH_SELECTION/e.g., On most days, <rephrase>the weather is warm and humid<rephrase>, with temperatures often soaring into the high 80s and low 90s Fahrenheit (around 31-34C).}

The output should be a markdown code snippet formatted in the following schema, including the leading and trailing "```json" and "```":
```json
{
    "response": string  // the selection to shorten
}
Don't forget to enclose the response value in double quotes.
\end{lstlisting}

Generation result example: On most days, the weather is warm
\newline
\newline

\textbf{Regenerate:}
\begin{lstlisting}[breaklines=true]
You will be provided with:
- A selected text
- A text containing that selection, with the selection marked using <regenerate> tags
- Your task is to substitute the selected text with a counterfactual example to diversify perspective, demographic, or approach. It should fit seamlessly into the original text. The regenerated selection must not disrupt the sentence structure or introduce grammatical errors (e.g., missing prepositions or incorrect tense).
- Examples: `toddler' changed to `adult', `terrorist' changed to `diplomat', `men' changed to `women', `easy' changed to `difficult', `great' changed to `poor'

Selection:
{TEXT_SELECTION/e.g., the weather is cold and icy}

Text with selection (wrapped in-between <regenerate> tags):
{TEXT_WITH_SELECTION/e.g., On most days, <regenerate>the weather is warm and humid<regenerate>, with temperatures often soaring into the high 80s and low 90s Fahrenheit (around 31-34C).}

The output should be a markdown code snippet formatted in the following schema, including the leading and trailing "```json" and "```":
```json
{
    "response": string  // the selection to regenerate
}
Don't forget to enclose the response value in double quotes.
\end{lstlisting}

Generation result example: On most days, the weather is cool and dry

\end{document}